\documentclass[12pt]{iopart}
\usepackage{graphicx, color}
\usepackage{hyperref}
\usepackage[utf8]{inputenc}

\expandafter\let\csname equation*\endcsname\relax
\expandafter\let\csname endequation*\endcsname\relax\usepackage{amsmath}
\usepackage{amssymb}

\newcommand{\al}{\alpha}
\newcommand{\be}{\beta}
\newcommand{\ga}{\gamma}
\newcommand{\de}{\delta}
\newcommand{\De}{\Delta}
\newcommand{\ep}{\epsilon}
\renewcommand{\th}{\theta}

\newcommand{\la}{\lambda}
\newcommand{\La}{\Lambda}
\newcommand{\Ga}{\Gamma}
\newcommand{\m}{\mu}
\newcommand{\n}{\nu}
\newcommand{\ka}{\kappa}
\newcommand{\na}{\nabla}

\newcommand{\si}{\sigma}

\newcommand{\om}{\omega}

\renewcommand{\d}{\text{d}}

\def\Noe{\mathcal{E}}
\def\UU{\mathcal{U}}

\def\ip{\hbox to4pt{\leaders\hrule height0.3pt\hfill}\vbox to8pt{\leaders\vrule width0.3pt\vfill}\kern 2pt}
 
\newcommand{\calC}{\mathcal{C}}
\newcommand{\calP}{\mathcal{P}}
\def\Aut{{\hbox{Aut}}}
\def\arr{\rightarrow}

\def\R{{{\mathbb R}}}
\def\GL{{\hbox{GL}}}

\newcommand{\pa}{\partial}

\newcommand{\beq}{\begin{equation}}
\newcommand{\eeq}{\end{equation}}

\def\ffrac[#1/#2]{\hbox{$\frac{#1}{#2}$}}
\def\Frac[#1/#2]{\frac{#1}{#2}}
\def\({\left(}
\def\){\right)}
\def\[{\left[}
\def\]{\right]}

\begin{document}

\title[Deflection light in conformal gravity]{Conformal gravity: light deflection revisited and the galactic rotation curve failure}

\author{M C Campigotto$^{1,2\footnote{Corresponding author}}$, A Diaferio$^{1,2}$ and L Fatibene$^{3,2}$}

\address{$^1$ Dipartimento di Fisica, Università di Torino,\\Via P. Giuria 1, 10125, Torino, Italy}
\address{$^2$ Istituto Nazionale di Fisica Nucleare (INFN), sezione di Torino,\\Via P. Giuria 1, 10125, Torino, Italy}
\address{$^3$ Dipartimento di Matematica, Università di Torino,\\Via C. Alberto 10, 10123, Torino, Italy}

\eads{\mailto{martacostanza.campigotto@to.infn.it}, \mailto{diaferio@ph.unito.it}, \mailto{lorenzo.fatibene@unito.it}}
\vspace{10pt}

\begin{abstract}
{We show how  Conformal Gravity (CG) has to satisfy a fine-tuning condition in order to be able to describe galactic rotation curves without the aid of dark matter, as
suggested in the literature}. If we interpret CG as a gauge natural theory, we can derive 
conservation laws and their associated superpotentials without ambiguities. We consider the light deflection of a point-like lens in CG and impose that the two Schwarzschild-like metrics with and without the lens at the origin of the reference frame are identical at infinite distances.
The conservation law implies that the free parameter $\gamma$ appearing in the linear term of the metric
has to vanish, { otherwise the two metrics are physically unaccessible from one other}. This linear term is responsible for mimicking the role of dark matter in the standard model and  it also appears in numerous previous investigations of gravitational lensing. Our result thus shows that the possibility of removing the presence of dark matter with CG { relies on a fine-tuning condition on the parameter $\gamma$}.

We also illustrate why
the results of previous investigations of gravitational lensing in CG largely disagree. These discrepancies derive from the erroneous use of the deflection angle definition adopted
in General Relativity, where the vacuum solution is asymptotically flat, unlike CG. In addition, the lens mass is identified
with various combinations of the metric parameters. However, these identifications are arbitrary, because the mass is not a conformally invariant quantity, unlike the conserved charge associated to the conservation laws.
Based on this conservation law {and by removing the fine-tuning condition on $\gamma$, i.e.~by setting $\gamma=0$}, the difference between the metric with the point-like lens and the metric without it defines a conformally invariant quantity that can in principle be used for (1) a proper derivation of light deflection in CG, and (2) the identification of the lens mass with a function of the parameters $\beta$ and $k$ of the Schwarzschild-like metric.
\end{abstract}

\submitto{\CQG}
%
\vspace{2pc}

\section{Introduction}
\label{sec:intro}
The last two decades have supplied cosmology with a great amount of observational data and consequently a phenomenological understanding of our Universe \cite{PlanckXIII2015}. However, our theoretical comprehension lacks a complete knowledge that simultaneously accounts for dark matter, dark energy and inflation, and that naturally fits into a quantum field theory, as particle physics does.
Nonetheless, the framework within which these data are usually interpreted is the assumption of the validity of general relativity (GR) which implies, as a consequence, the $\La$-Cold Dark Matter ($\La$CDM) model as the currently standard cosmological model. GR is known as the most accredited and elegant theory of the gravitational force. It has been tested at the Solar system scale \cite{TestGRJainKhoury,TestGR2015} and recently the first detections of gravitational waves by binary systems were a further brilliant confirmation of a revolutionary prediction of GR \cite{GW150914,LVT151012,GW151226,GW170104,GW170814,GW170817}.
Still, GR describes the dynamics of cosmic structures and the expansion history of the Universe only if we suppose the existence of dark matter and dark energy. However, the dark matter particles remain unidentified and the nature of dark energy is still unknown. In addition, the $\La$CDM model, with its six parameters, is not a definitive model: it presents a number of problems on very non-linear scales \cite{DelPopolo2017}.
An alternative approach is to describe the phenomenology of cosmic structure without dark components by focusing on the left hand side of Einstein equations and modifying the gravity theory. 

Conformal gravity (CG) has been proposed as an alternative theory of gravity where an additional invariance principle is imposed \cite{Mannheim:1989}. The additional symmetry is a local conformal invariance that requires the action to remain invariant under any and all local {\it conformal {(Weyl)} transformations} of the metric 
\beq
g_{\m\n}(x) \mapsto \Phi(x) g_{\m\n}(x)\>,
\label{ConfTransf}
\eeq
where $\Phi(x)$ is an arbitrary regular and positive function of the spacetime coordinates $x$.
{Let us stress that here these conformal transformations act on the metric field, leaving the position $x$ on spacetime unaffected.}
The conformal invariance provides a traceless stress-energy tensor and a vanishing Noether current associated to it, showing that conformal transformations are pure gauge and thus non-dynamical \cite{Campigotto2014,Jackiw,SuatDengiz2016}.

The {most general} exact static and spherically symmetric vacuum solution of the field equations of CG has been derived in \cite{Mannheim:1989,Mannheim:newton}:
\beq
\begin{aligned}
ds^2 =&\Phi(\mathbf{r})\left[-A(r) dt^2 + \Frac[1/A(r)] dr^2 + r^2 (d \th^2 + \sin^2\th d\phi^2)\right]\; , \\
A(r)=& 1- \frac{\be(2-3\be\ga)}{r} -3\be\ga+\ga r - k r^2 \; .
\end{aligned}
\label{solutionCG}
\eeq
The metric exhibits two new extra terms, parametrized by $\ga$ and $k$, in addition to the constant $-3\be\ga$ and the standard Schwarzschild terms, parametrized by $\be(2-3\be\ga)$. Galactic and cosmological observations were used to constrain these parameters \cite{Mannheim:1989,Pireaux:2004,Pireaux:2004b,Diaferio2011}.

Let us remark that in \cite{Mannheim:2005} (see equation (202) there) a different, more general, expression is found for the solution. However, that expression is determined by solving a linear combination of field equations, which is not equivalent to the original field equations, there denoted by $W_{\mu\nu}=0$. If one further imposes the field equations, we eventually find the metric (\ref{solutionCG}), that thus is the most general 
static and spherically symmetric vacuum solution of CG. The constant contribution $1-3\be\ga$ needs to be in that form. 
Any modification of it, e.g.~to $1+ w-3\be\ga$ shows that for the metric (\ref{solutionCG}) to be a solution one needs to set $w=0$ or $w=-2+6\be\ga$, which, however, is excluded by considerations about the signature. 
That form of the constant contribution will be essential for us below.

CG has been presented as a field theory with the extra gauge group of conformal Weyl transformations. This feature suggests that CG can in principle be renormalisable.
However, when, in previous work, applications to astrophysics or cosmology are considered, the gauge invariance is arbitrarly broken with the aim to fit the data, which are definitely not conformally invariant. Unfortunately, the mechanism of symmetry breaking does not appear to have been explicitly described in the literature.

This issue clearly is fundamental for the definition of the {\it mass} generating the gravitational field. For example, in the static spherically symmetric solution (\ref{solutionCG}), we have at least two different options: either we search for conserved quantities associated to the solution, or we attempt to derive the Newtonian limit and define the mass through the Kepler laws, which will in general be approximate. The latter approach is usually used in astrophysics.
In this case, however, a solution comes with all the metrics which are conformally equivalent to the solution itself. Each of these metrics describes test particles, with their own mass, which move
along different timelike geodesics; these geodesics, on turn, depend on the conformal factor $\Phi$. In order to fit the data, one can arbitrarily choose a form for $\Phi$, and break the conformal invariance, as it has been done in the literature \cite{Mannheim:1989};  for that specific metric, one can derive an expression for the central mass $m$ generating the field. Unfortunately, this appraoch is dangerously similar to define the electric charge as a gauge non-invariant quantity.

Although one could argue that these choices are {\it physically sound}, to legitimately adopt this approach, one should (1) clearly state that these choices are hypotheses which are independent of the setting of CG as a field theory, (ii) argue that they are not contradictory, namely that mutually exclusive different hypotheses are not invoked in different contexts, as {\it ad hoc} hypotheses to fit the data, and (iii) provide a specific and detailed interpretation of the theory, for example by showing how the symmetry is broken, if it is indeed broken.
To the best of our knowledge, these clarifications are not present in the literature and it thus appears hard to 
currently define a single CG theory that is mathematically sound.

In \cite{Mannheim:1995gal,Mannheim:2010fit} and references therein, the authors show that, by setting $\Phi(\mathbf{r})=1$, CG is able to reproduce the galactic rotation curves of a sample of 111 disk galaxies without adding any dark matter component. Their analysis finds that, by associating $\beta$ to the galactic baryonic mass, $\ga$ and $k$ are universal constants. Therefore, departures of galactic rotation curves from the Newtonian expectation based on the distribution of the luminous matter alone can be entirely of cosmological origin, encoded in the values of the parameters $\ga \sim 3.06 \times 10^{-30} cm^{-1}$ and $k\sim 9.54 \times 10^{-54} cm^{-2}$ \cite{Mannheim:2010fit}. In passing, we remind that Horne \cite{Horne2016} has recently shown that this interpretation might however be too simplistic when a conformally coupled Higgs field on CG is taken into account.

In CG, both the cosmological constant problem and the flatness problem are naturally solved \cite{Mannheim2017}: CG indeed satisfactorily describes the observed Hubble diagram of supernovae and gamma-ray bursts, similarly to $\Lambda$CDM, but without requiring the existence of dark energy \cite{Diaferio2011}. Moreover, in \cite{Mannheim:1989lambda}, Mannheim claims that the quadratic term $k r^2$ may be associated to a de Sitter background geometry. In fact, the de Sitter metric is a vacuum solution of CG, even if CG does not contain any {\it ad hoc} cosmological term in the action. The drawback of this feature, however, is the fact that CG is unable to reproduce the observed abundance of deuterium \cite{Elizondo1994}: the Universe expansion is always accelerating and the expansion during the phase of the cosmological nucleosynthesis is thus slower in CG than in the standard model, implying a longer phase of deuterium burning. 

Despite this drawback, CG has additional attractive features, like its renormalizability \cite{Mannheim:2011Making} and the unnecessity of an initial Big Bang singularity. In fact, conformal invariance has been proven to be crucial to removing all kinds of spacetime singularities \cite{thooft2009,Modesto2016}. Moreover, in CG, when a matter action is taken into account, we can derive a conformal cosmology in which gravity is globally repulsive rather than attractive \cite{Mannheim1999}.

An additional important topic is the investigation of the formation of the large-scale cosmic structure from small initial density perturbations; unfortunately, the work on this topic in the literature \cite{Mannheim:2011pert,R2} is still too limited to enable the drawing of definitive conclusions. These studies would in principle be relevant to explore the possible alleviation of the serious discrepancy between the observed thermal properties of $X$-ray clusters and the CG expectations \cite{Horne2006,Diaferio2009}. For the sake of completeness, we finally mention that solutions corresponding to gravitational waves have also been investigated in \cite{Myung:2014cra,Fabbri:2011zz}.

In this work, we focus on the study of the {point-like source} solution \eqref{solutionCG} of CG and its phenomenological consequences on gravitational lensing and the dynamics of disk galaxies. We {argue that} conservation laws imply  that the solution  \eqref{solutionCG} reduces to a Schwarzschild-de Sitter-like metric with $\ga=0$. 

{
To be more precise, we are not arguing that $\ga\not = 0$ are not solutions in CG. 
We are arguing that, in view of conservation laws, these solutions are not {\it physically accessible} one another as defined below, unless one declares not to be interested in conservation laws.
We shall in fact show that if one tries to define conserved quantities by the most general and liberal framework available, then a metric (\ref{solutionCG}) is isolated in the space of solutions, i.e.~one cannot define the conserved quantities of it relative to any other solution.

That claim does not rely on identification of the conserved quantity with the physical energy, so extra care is in order.}
This result obviously {would frustrate} the attempts suggested in the literature to describe both the observed gravitational lensing phenomena and the velocity rotation curves of disk galaxies with a correction term depending on $\ga\not=0$.

In Section \ref{introCG}, we review how CG can be described in the framework of gauge natural theories. 
In Section \ref{sec:conservations}, we derive the relative conservation law. 
In Section \ref{sec:gamma}, we discuss how this conservation law implies $\gamma=0$, and in Section \ref{sec:rotation} we {point out 
various inconsistencies on light deflection in CG which are present in the literature and discuss how our result clarifies them}.

Section \ref{sec:rotation} is  kept independent of the argument about conservation laws, and it  cannot thus be fully rigorous. 
We think that it is worth noticing that $\ga=0$ implies a failure of fitting the phenomenology of the deflection of light.
A number of issues and ambiguities in the  literature require a clarification independently of the conservation law argument.
The reverse argument also holds: if we want to preserve the interesting features that CG might have with $\ga\not=0$, we need to provide a still unknown recipe for the conservation laws that does not imply $\ga=0$, as instead we find here with the currently known recipes.

\section{Conformal Gravity as a Natural and a Gauge Natural Theory}
\label{introCG}

The {vacuum} action of CG is given by  
\beq
\begin{aligned}
S_W=& - \ka  \int \! d^4x \left( -g\right)^{1/2} C_{\la\mu\nu\ka}C^{\la\mu\nu\ka}=\\
=&- \ka  \int \! d^4x \left( -g\right)^{1/2} \( R^2 - 6 R_{\mu\nu}R^{\mu\nu}+ 3 R_{\la\mu\nu\ka}R^{\la\mu\nu\ka}\)
\end{aligned}
\label{CGaction}
\eeq
where $C_{\la\m\nu\ka}$ is the Weyl tensor, $g$ is the determinant of the metric and $\ka$ is a {coupling} constant \cite{Mannheim:1989}. The action \eqref{CGaction} is the unique combination 
of four-dimensional diffeomorphism invariants \cite{Campigotto2014} which is also invariant under the local conformal invariance (\ref{ConfTransf}).
It is the gauge transformation that leaves the theory invariant. Under the gauge transformation \eqref{ConfTransf}, the Weyl tensor transforms as
\beq
C_{\la\m\n\kappa} \mapsto \Phi(x) C_{\la\m\n\kappa} \; ,
\eeq
while the Ricci and Riemann tensors, that are covariant under any change of spacetime coordinates, transform with a combination of derivatives of $\Phi(x)$.
Consequently,  the Lagrangian density in a four-dimensional spacetime is conformally invariant.

{
In the literature sometimes one finds a different action for CG, obtained by adding a Gauss-Bonnet term $3G$, where
\beq
G=  \ka \int \! d^4x \left( -g\right)^{1/2} \( R^2 - 4 R_{\mu\nu}R^{\mu\nu}+  R_{\la\mu\nu\ka}R^{\la\mu\nu\ka}\)
\eeq
 which, in four dimensions  is  a global divergenceless term with a local potential, so that it does not affect the field equations;
however, this term is not conformally invariant and it thus spoils the conformal invariance. 
Moreover, adding a divergence to the action functional usually does affect the conservation laws, potentially introducing divergences, as we shall argue below.
For these reasons, we avoid it. 
Anyway, when dealing with conserved quantities, we shall consider  the action functional $S=\al S_W + \la G$, so that all the situations can be obtained by specifying $\la$
and we can directly prove what depends on the Gauss-Bonnet term and what does not; {similarly, we
can discuss} how the asymptotic behaviour of the conserved quantities is affected by the Gauss-Bonnet term, i.e.~by $\la$.

\medskip

In a recent work, Campigotto and Fatibene presented CG both as a gauge natural theory \cite{Campigotto2014} and a natural theory \cite{Campigotto2015,Campigotto2016}. We refer to \cite{FatibeneLibro, Kolar, Eck} for the general notation and framework.
We {review} the basic framework of their analysis in Appendix $A$. 

{
By fitting CG into the framework of gauge natural theories (see also \cite{Campigotto2014,Jackiw,SuatDengiz2016}), we get a canonical treatment of conservation laws for free and in particular a theorem ensuring that all Noether currents are
always exact forms. In addition, we have full control of the global properties of the fields and their observability. 
This framework has proven to be suitable to discuss gauge theories in their generally relativistic formulations, as well as couplings between spinor fields and gravity (see \cite{Spinors}).
Thus conformal gravity comes with a structure to compute conserved quantities.
We illustrate this point in the Section below.

}

\section{Symmetries, Superpotential, and Conserved Quantities}
\label{sec:conservations}

{
In gravitational theories, conserved quantities are hardly ever obtained by simply integrating Noether currents over a spatial volume. 
It is very common, even in standard situations, that such integrals diverge or give the wrong results (see \cite{{Katz:1985}}). 
If, as in this case, one has point-like sources, then Noether currents diverge, as they diverge if one considers unbound regions.
Both these issues are present in our case and they lead to improper integrals over a non-compact region, which are not {\it a priori} finite.

For avoiding infinities, one first shows that Noether currents are exact forms, then uses Stokes theorem to reduce conserved quantities to integrals over closed surfaces 
which can easily avoid point-like sources and their infinities. However, sooner or later one should also send this surface to infinity (here by taking the limit of the radius of the sphere to infinity)
and there, generically, infinities are back, unless one requires asymptotically flat fields or introduces counterterms in the action.

Since solutions (\ref{solutionCG}) are not asymptotically flat, one needs countertems as a more general recipe.
In \cite{FatibeneAugmented2005} we introduced a framework for conserved quantities in which one can compute the conserved quantity relative to a {\it reference configuration}, 
without relying on asymptotic flatness, solving anomalous factors in many standard cases {(see \cite{Kerr:1999, BTZ:1999, TaubBolt:2000, TaubBolt2:2003, Bibb:2009})}, 
maintaining general covariance, and being independent of addition of divergences to the action.

The framework essentially selects a counterterm so that the conserved quantity corresponds to a deformation of the solution to a chosen reference configuration.
The selected boundary terms compensate in case one adds a divergence to the Lagrangian so that the conserved quantity turns out to be unaffected.
The price paid, is that the solution and the reference configuration have to agree on the boundary.

This, to the best of our knowledge, is the most general and liberal framework for conserved quantities, which in fact reproduces ADM, boundary terms, pseudo-tensors in the situations 
when these apply {(see \cite{BY:2001, CS:2006, ADM:2012})}.

So we say that two solutions are {\it physically accessible} one to the other if they induce the same metric at infinity, so that conserved quantities can be defined, they are guaranteed to be finite,
they are independent of the divergences that may be added to the action. As we shall show in the next Section, this is what will allow to argue that either $\ga=0$ or the solution is not physically accessible.

}

Let us consider the action functional $S=\al S_W + \la G$, including for the sake of discussion a Gauss-Bonnet term which does not affect field equations, and the infinitesimal transformation, $\xi= \xi^\ep \pa_\ep$ {(which defines global  transformations on fields only in view of the discussion about gauge natural structures in Appendix A)}; the corresponding Noether current is the $3$-form $\Noe$,
\beq
\Noe=\left(T^\la{}_\ep \xi^\ep + T^{\la\mu}{}_\ep \na_\mu \xi^\ep + T^{\la\mu\nu}{}_\ep \na_{\mu\nu} \xi^\ep\right) d\si_\la,
\eeq
where $\d \si_\lambda=\sqrt{g} \ep_{\la \mu_1\mu_2 \mu_3}\d x^{\mu_1}\wedge\d x^{\mu_2}\wedge \d x^{\mu_3}$ is the local basis of the $3$-forms,
and
\beq
\begin{aligned}
T^\la{}_\ep=& 3(\la+\al) RR^\la_\ep +4(3\al+2\la) R^{\mu\nu} R^\la{}_{\mu\nu\ep} - 2(3\al+2\la) R^{\la\nu} R_{\ep\nu} -L\de^\la_\ep\\
T^{\la\al}{}_\ep=& -4(\la+\al) \na^{(\al} R_\ep{}^{\la)}
-2(\la+\al) \na_\ep R g^{\al\la} 
+2\la  \na^\la R \de^\al_\ep
-12\al   \na^\la R^\al_\ep +\\
&+4\la   \na_\ep R^{\la\al}
+4\la  \na^\al R_\ep{}^{\la}
\\
T^{\la\mu\nu}{}_\ep=& 2(\la+\al)Rg^{\mu\nu}\de^\la_\ep -2(\la+\al)R g^{\la(\mu}\de^{\nu)}_\ep-4(3\al+2\la) R^{\mu\nu} \de^\la_\ep+\\
& +4(3\al+2\la) R^{\la(\mu}\de^{\nu)}_\ep - 4(\la+3\al) R_\ep{}^{(\mu\nu)\la}.
\end{aligned}
\eeq
where $L$ is the Lagrangian density for the action $S$, namely
\beq
L= \sqrt{ -g} \[ (\al+\la) R^2 -(4\la+ 6\al) R_{\mu\nu}R^{\mu\nu}+ (3\al+\la) R_{\la\mu\nu\ka}R^{\la\mu\nu\ka}\] \, .
\eeq

{
	The Noether current, as usual, is obtained just by a wisely chosen sequence of covariant integrations by parts. Some details can be found in \cite{FatibeneLibro}.
}

The Noether current $\Noe$ can be recast as $\Noe=\tilde{\Noe} + \textup{div} \, \mathcal{U}$, as already anticipated.
We can see that $\Noe$, and thus the superpotential $\mathcal{U}$, only depends on the spacetime transformations, i.e.~the infinitesimal generator of diffeomorphisms, and does not depend on conformal transformations, i.e.~the gauge transformations of the metric.

The superpotential is the $2$-form
\beq
\UU=\frac{1}{2} \left\{ \left( T^{[\la\mu]}{}_\ep -\frac{2}{3} \na_\nu T^{[\la\mu]\nu}{}_\ep \right)\xi^\ep+ \frac{4}{3} T^{[\la\mu]\nu}{}_\ep \na_\nu \xi^\ep \right\}\d \si_{\la\mu}
\eeq

{
As required by {\it augmented variational principles (AVP)} \cite{FatibeneAugmented2005}, we fix a reference metric $\bar{g}$ 
\beq
A_{\bar{g}}(r)=1-\frac{\bar{\be}(2-3\bar{\be}\bar{\ga})}{r}-3\bar{\be}\bar{\ga}+\bar{\ga} r-\bar{k}r^2 \; ,
\eeq
and the relative energy is $Q= \int_S (\UU -\bar \UU -  \xi^{\al} \De^{\la} d\si_{\la\al} )$
where the integral is computed on the sphere $S$ at infinity, $\bar \UU$ denotes the superpotential evaluated along the reference metric $\bar g$, and $\De=\De^\la  d\si_\la$ is the $3$-form, acting as a counterterm, defined as
\beq 
\De^\la =  2(\al+\la) R g^{\mu\nu} w^\la_{\mu\nu} - 4(3\al + 2\la) R^{\mu\nu} w^\la_{\mu\nu} - 4(3\al+\la) R_\al{}^{\be\nu}\la q^\al_{\be\nu}
\eeq
Here we set $w^\la_{\mu\nu} = u^\la_{\mu\nu}- \bar u^\la_{\mu\nu}$, $u^\la_{\mu\nu}= \Ga^\la_{\mu\nu} - \de^\la_{(\mu} \Ga^\al_{\nu)\al}$, $\bar u^\la_{\mu\nu}$ is the corresponding quantity computed along the reference metric, $q^\la_{\mu\nu} = \Ga^\la_{\mu\nu} - \bar \Ga^\la_{\mu\nu}$, $\Ga$ and $\bar \Ga$ are the Christoffell symbols of $g$ and $\bar g$, respectively.
Accordingly, the form $\De$ is a global form on spacetime, since both $w^\la_{\mu\nu}$ and $q^\la_{\mu\nu} $ are tensors.

The quantity $Q$ defined for the vector $\xi=\partial_t$ is called the {\it conformal mass}. It is conformally invariant and it globally depends on the metric and the reference background.

Let us first consider $\al=1$, $\la=0$, namely the Weyl action without the Gauss-Bonnet correction, and $g$ and $\bar g$ just imposing $k=\bar k$ and $\ga=\bar \ga=0$. The conformal energy $Q$ turns out to be
\beq
Q= -24(\be-\bar\be)k .
\label{consQ}
\eeq

On the other hand, if we simply consider the integral of superpotential for the solution, for the general action functional $S=\al S_W + \la G$, i.e.~including the Gauss-Bonnet correction depending on $\la$,  
we have 
\beq
\begin{aligned}
\hat Q=&-2\la (2 k^2  r^2 -3 \ga k  r + \ga  (\ga+6 k \be)  )r +\\
+&2 \be (3 \ga ( \be k (\la +6  \al)- \ga (\al- \la) ) - 12 k \al-2 k \la ) 
 +\frac{18 \be^2 \ga^2 \al}{r}
 + O(r^{-2}) .
\end{aligned}
\label{consQ2}
\eeq

$\hat Q$ diverges for $r\arr+\infty$, unless one sets $\la=0$ (which corresponds to no Gauss-Bonnet term, which is the prescription in \cite{FatibeneAugmented2005}) or $\ga=0$ (and $k=0$).
 In passing, we note that $\la=-3\al$ is quite an usual choice in the current literature.
Thus, when it makes sense, the simple integral of the superpotential specialises to the AVP prescription.

Therefore, in analogy with standard GR where the same conserved quantity associated to $\xi=\partial_t$ is known to provide a definition of the mass source, our equation \eqref{consQ} suggests that there might be a link between the mass of the source of the gravitational
field in CG, which is not a conformal invariant quantity, and the conformal invariant charge $Q$.  

We can define such a non-conformally invariant mass, by imposing a local Newtonian limit to the adopted metric. For the Schwarzschild-like metric (\ref{solutionCG}), where, as done for the galaxies, the conformal invariance has been arbitrarily broken by choosing $\Phi=1$ \cite{Mannheim:1989}, we impose the condition that a test particle moves around a mass $m$ accordingly to the Newtonian expectation and its corrections in GR.
Consistently, although CG does not actually have a Newtonian limit  (see \cite{Mx}), we
should still expect that the GR Schwarzschild solution is comparable to the CG solution (\ref{solutionCG}) in a region of spacetime which is not too far away, as we specify below, from the source. Although a rigorous analysis would clearly be necessary here, we can still expect that this argument holds, at least in this specific situation, because the motion of a test particle is determined by the metric in a small spacetime volume around the particle, and should be indepent of the asymptotic behaviour of the metric itself.
By adopting this argument, we can derive a definition of {\it Newtonian local mass} that is independent of the conformal mass obtained from conservation laws. 

In practice, we can simply require that, in equation (\ref{solutionCG}), $A(r=2m)=0$ and $A'(r=2m)= \frac{1}{2m}$, as in the GR Schwarzschild solution with a field source of Newtonian mass $m$. In regions of the spacetime close to the source, where $k$, in the solution (\ref{solutionCG}), is small enough to be irrelevant, the test particle moves accordingly to the Newtonian law, as it happens in the Solar System and in all the kinematic phenomenology investigated in CG, used to constrain the value of $\beta$, $\ga$, and $m$.

By imposing the requirements  $A(r=2m)=0$ and $A'(r=2m)= \frac{1}{2m}$, we find that there are two sets of parameters, namely $\{\be=m$, $\ga=0\}$ and $\{\be\simeq \frac{4m}{3}$, $\ga\simeq \frac{1}{2m}\}$, which approximate the GR Schwarzschild solution, hence the Newtonian solution, for a source of mass $m$.

We thus find that the first solution again has $\ga=0$, as we argue based on conservation laws, whereas the second solution yields the mass parameter
$m= \be \(1-\frac{3}{8}\be\ga\)$, which, to the best of our knowledge, has never been considered as a candidate to describe the Newtonian local mass.

Our result also shows that the identification $m= \be \(1-3\be\ga\)$, that has
been used in the literature to identify the GR Schwarzschild solution with the CG Schwarzschild-like solution, does not actually qualify as an appropriate choice.
Below, we discuss the consequences of our result in the phenomenology of gravitational lensing.

In GR, the absence of the central source yields the Minkowski metric, or the AdS metric if we include a non-null cosmological constant $k$, as the background metric. Unfortunately the background metric
is unknown in CG, because of the ambiguity of the identification of the metric parameters with the source's physical local mass. This crucial point has never been properly addressed in the literature as we do here, and it is the origin of most of the discrepancies among the investigations on the gravitational lensing phenomenology in CG. 

In the following Section, we illustrate how looking for a zero point of the superpotential is a possible approach for a proper, formal 
definition of a background metric. However, this argument leads to derive that we must necessarily have $\ga=0$.

\section{Why $\ga=0${?}}
\label{sec:gamma}

{In order to highlight the differences with conformal gravity solutions, we can discuss the situation of a monopole metric (see \cite{Barriola:1989}) defined by
\beq
A(r) = 1-8\pi G\eta^2  -\frac{2MG}{r}
\label{monopole}
\eeq
which of course, for any fixed non-zero $\eta$, is not asymptotic to Minkowski  metric.
It is instead {\it asymptotically locally flat} (ALF); see \cite{Barriola:1989} and \cite{ALF}.
For that reason, it is not accessible from Minkowski.\footnote{The metric (\ref{monopole}) cannot reduce to a Minkowski metric at large $r$ with a redefinition of the coordinates: in fact, $r$ cannot be redefined because it appears in front of the angular part, and there is no redefinition of $t$ that can eliminate the constant $\eta$ from $A(r\to\infty)=1-8\pi G\eta^2$.}
If one looks for accessible metrics which define the same metric at infinity, one can consider, for example,
$A(r) = 1-8\pi G\bar\eta^2  -\frac{2\bar MG}{r}$, which is accessible if and only if $\eta=\bar \eta$.
We have no condition on $M$, meaning that the metric for $(M, \eta)$ is accessible from $(\bar M, \eta)$, and from $(\bar M=0, \eta)$ in particular.

That means that the monopole metric cannot be produced in a classical regime; however, if a monopole is there from the beginning,
it can capture mass to produce a monopole black hole as in (\ref{monopole}).

This agrees and proves the interpretation of the monopole solution presented in \cite{Barriola:1989}.
However, it is not what happens in CG. Here, we have three terms threatening accessibility (namely the constant, the ones in $r$ and in $r^2$) parameterized by $(k, \ga, \be)$.
If we want to get rid of these terms, one needs either to fix  all the three parameters or fixing  $k$ and $\ga=\ga'=0$ leaving $\be$ unconstrained.

In the first case, all the parameters are frozen meaning that the solution with $\ga\not= 0$ is not classically accessible from any other solution; if it is there, it is there from the beginning and it cannot change at all,
not even by capturing mass around.
In the second case, when one sets $\ga=0$, the solution is accessible from analogous metrics with different values of $\be$ and the relative conformal energy (with respect to $\bar \be=0$) can be computed in AVP framework to be  $Q=-24\be k$.

}

We thus have
\beq
A_{g}(r)=1-\frac{2\be}{r}-kr^2, 
\qquad\qquad 
A_{\bar{g}}(r)=1-\frac{2\bar{\be}}{r}-kr^2.
\label{physicalmetric}
\eeq
and the corresponding relative conformal energy is 
\beq
Q= -24k(\be-\bar \be)
\label{masscorrect}
\eeq
We can see that the charge needed to move from the vacuum configuration to the configuration of the field source 
does only depend on $\be$ and $k$. 

We conclude that the various identifications of the mass term in the metric solution proposed in the literature (e.g., Mannheim and Kazanas \cite{Mannheim:1989}) is not supported by our
rigorous approach within the framework of gauge natural theories. Equation \eqref{masscorrect} suggests that a quantity proportional to the product $k\beta$ is 
a more appropriate identification of the source local mass. 

In \cite{Mannheim:1995gal,Mannheim:2010fit}, the $\ga$ term is essential to the good fitting of the rotational curves of several disk galaxies. Unfortunately, our constraint $\ga=0$  poses serious challenges to CG, because, once the metric \eqref{solutionCG} reduces to a Schwarzschild-de Sitter spacetime, it is unable, exactly like GR, to reproduce the galactic phenomenology without the aid of a dark matter component. 

\section{Consequences on Light Deflection in CG}
\label{sec:rotation}

Given our result, we provide here a brief review of the investigations of light deflection in CG presented in the literature. The most relevant issue of previous work is the severe disagreement between different studies. The disagreement concerns (1) the sign of $\ga$ \cite{Edery:1997,Pireaux:2004}; (2) the association of the mass of the lens to different combinations of the parameters in the metric \eqref{solutionCG}\cite{CattaniScalia2013,Sultana:2013}; (3) the choice of the geometric definition of the deflection angle \cite{Rindler:2007}.

Firstly, Edery and Paranjape \cite{Edery:1997} derived the total deflection angle for a point-like lens with the standard formula for the asymptotically flat metric solution given by Weinberg \cite{Weinberg}. They recover that, in order to obtain a stronger deflection than the GR prediction, thus avoiding the requirement of dark matter, the sign of $\ga$ has to be negative, in contrast with the positive value found by fitting galactic rotation curves \cite{Mannheim:1995gal,Mannheim:2010fit}. Moreover, in this treatment, the deflection angle increases linearly with the impact parameter. At odds with their results, Pireaux \cite{Pireaux:2004,Pireaux:2004b} found that the sign of $\ga$ depends on the nature of the particles considered: $\ga <0$ for photons or relativistic particles, while $\ga >0$ for massive or non relativistic particles. Both approaches \cite{Pireaux:2004,Edery:1997} identify the mass of the lens with the parameter $\be$. These apparently inconsistent results clearly derive from the incorrect assumptions of a Minkowski metric at large distances from the lens and by the unjustified identification of the lens mass with $\be$.

As illustrated in the previous Section, the vacuum solution proposed by Mannheim and Kazanas \cite{Mannheim:1989} is not asymptotically flat; therefore, the standard total deflection angle formula, used by \cite{Edery:1997} and \cite{Pireaux:2004} is inappropriate for CG and should be replaced by a more general recipe. 

Rindler and Ishak \cite{Rindler:2007} proposed an alternative definition of the total deflection angle that can be applied to geometries that are not necessarily flat at large distances from the lens. They resort to the invariant formula for the cosine of the angle $\psi$ between two coordinate directions, $d^i$ and $\de^i$ 
\beq
\cos \psi= \frac{g_{ij}d^i \de^i}{\left(g_{ij} d^i d^j\right)^{1/2}\left(g_{ij} \de^i \de^j\right)^{1/2}} \, ,
\eeq
where $g_{ij}$ is the 2-metric in the spatial equatorial coordinate plane, where the longitudinal angle $\th=\pi/2$.
For the first time, they show, in the GR framework, that in a Schwarzschild-de Sitter spacetime, where the cosmological constant $\La$ is not null, there is a contribution of $\La$ to the deflection angle. In previous work, this contribution is  neglected, because $\La$  drops out of the differential equation for a light path and the Schwarzschild metric is used in place of the correct Schwarzschild-de Sitter metric. In GR, the contribution of $\La$ to the bending of light is of the order of $\La R^3c^2/GM$, where $M$ is the lens mass and $R$ is linked to the distance of closest approach of light; it is thus relatively smaller than the leading term $2GM/c^2R$. Neglecting the contribution of $\La$ might be harmless in practical measurements, but, in principle, it is rigorously incorrect.  

Sultana and Kazanas \cite{SultanaKazanas,Sultana:2013} and Cattani et al.~\cite{CattaniScalia2013} adopt the Rindler-Ishak approach to estimate 
the deflection angle in CG with the Schwarzschild-like metric \eqref{solutionCG}. However, they arbitrarily associate the lens mass to two different combinations
of the metric paramaters: Sultana and Kazanas  \cite{SultanaKazanas,Sultana:2013} adopt $M=\be$, whereas Cattani et al. \cite{CattaniScalia2013} adopt $M=\be(2-3\be\ga)/2$. Thanks to their definition for the lens mass, Cattani et al. find that a positive  $\ga$, as required by the rotational curves of disk galaxies \cite{Mannheim:1995gal,Mannheim:2010fit}, increases the deflection angle. On the contrary, Sultana and Kazanas find that a positive  $\ga$ decreases the deflection angle. 
At the same time, however, Sultana and Kazanas \cite{SultanaKazanas,Sultana:2013} find that the contribution to the deflection angle 
of the cosmological term proportional to the parameter $k$ of the Schwarzschild-like metric \eqref{solutionCG} 
dominates over the term proportional to $\ga$; therefore, they conclude that
the fact that this latter term decreases the deflection angle appears to be irrelevant in the CG lensing phenomenology. However, our result illustrated in the previous Section indicates that this conclusion turns out to be irrelevant, as far as it is based on the arbitrary assumption $M=\be$.  
Similarly, the attempt of Lin and Wang \cite{LimWang2016} of ascribing the inconsistent results of  \cite{SultanaKazanas,Sultana:2013} and \cite{CattaniScalia2013} to the order of the approximation in $M$ and $\ga$ of the deflection angle expression is not supported by any physical argument.
This statement is confirmed by the exact analytical solution to the null geodesic in the Schwarzschild-like metric \eqref{solutionCG} with $M=\be$ \cite{Villanueva2013}. This solution,  involving the p-Weierstrass elliptic function, is fully consistent with the approximate solution of \cite{SultanaKazanas} and \cite{Sultana:2013} and demonstrates that the problem is not in the order of the approximation.

In conclusion, the description of the phenomenology of light deflection in 
CG currently present in the literature is still inconclusive. The correct description 
requires the use of the appropriate behaviour of the metric at large distances from the lens, as adopted in the approach of Rindler and Ishak \cite{Rindler:2007},  and the identification of the lens local mass with the appropriate combination of the parameters in the Schwarzschild-like metric \eqref{solutionCG}. Our description of CG as a gauge natural theory implies that $\ga=0$ and that the lens mass should be associated to some function of the product of the parameters $\be$ and $k$ (see equation \eqref{masscorrect}). It remains to be seen whether this analysis is worth to pursue further, because, being $\ga=0$, CG looses most, if not all, of its astrophysical appeal. 

{
Let us also mention that the physical mass of the source is defined through the theory's Newtonian limit, essentially using Kepler third law, which is definitely not conformally invariant.
On the other hand the conserved quantity $Q$ is conformally invariant so the relation between the two quantities needs to be further clarified.
}

 Our findings suggest an intriguing interpretation. We have two masses: the first mass, the local mass, is set by the central field source; the second mass, the conformal mass, is asymptotic in nature and is set by both the local source and the cosmological constant $k$.

\section{Conclusions}

We show that the $\ga$ parameter in the metric solution of CG proposed by Mannheim and Kazanas \cite{Mannheim:1989} is fixed to zero by conservation laws. Our result is a consequence of our demonstration that CG is a gauge natural theory and we can thus apply a standard treatment of the conserved quantities associated to the symmetries of the theory. CG has two global symmetries: the conformal symmetry and the diffeomorphism symmetry. The conserved charge corresponding to the conformal symmetry is zero; in other words, this symmetry has no dynamical role. For the identification of the conserved charge associated to the diffeomorphism symmetry, we study the asymptotic behaviour of two metric solutions corresponding to the background vacuum and to a single massive source of the gravitational field. We identify the conserved charge with the conformal energy required to move from one solution to the other; we derive this energy as the difference of the conserved charges associated to the time component of diffeomorphisms. We find that the conserved charge is proportional to the product of the parameters $\be$ and $k$, and we obtain the necessary condition $\ga=0$.

We also find that $\ga=0$ clearly satisfies the requirement that a test particle motion in a small spacetime volume close to the source has a Newtonian behaviour. The same Newtonian behaviour is also obtained with the solution $\{\be\simeq \frac{4m}{3}, \ga\simeq \frac{1}{2m}\}$, which leads to the identification of the mass $m= \be \(1-\frac{3}{8}\be\ga\)$. We thus conclude that none of the mass identifications presented in the
literature does actually reproduce the Newtonian motion of a test particle in spacetime regions where the constant $k$ is negligible, like the Solar System case. Unfortunately, however, the identification $m= \be \(1-\frac{3}{8}\be\ga\)$ is incompatible with the conservation laws and only the solution $\ga=0$ appears to remain viable.

Our result implies that the parameters of the Mannheim and Kazanas metric solution may not be trivially related to the physical quantities, like the lens local mass, involved in the gravitational lensing effect. We show how those identifications of various combinations of the parameters with the lens mass suggested in the literature are not supported by any physical argument and they thus lead to inconsistent results.

The $\ga$ parameter is responsible for reproducing the galactic rotation curves in CG without the aid of a dark matter component: this ability is one of the main reasons why CG gained appeal.
The constraint we derive proves that  the current formulations of  CG can unfortunately describe the phenomenology of neither galactic rotation curves nor gravitational lensing. 
With our result, the metric solution reduces to the same geometry of a Schwarzschild-de Sitter spacetime in GR; therefore, the contribution of $\be$ and $k$ to the deflection angle is analogue to the contribution of the usual mass term $M$ and the cosmological constant $\La$. In GR, the latter contributes with the term $\La R^3 c^2/ GM$, which is suppressed compared to the leading term $GM/c^2R$.

 We finally wish to emphasize that that conservation laws are related to CG as a field theory. On the contrary, the arguments about dark matter and dark energy  are related to the solutions rather than to the setting of the field theory. We certainly agree that the solution (\ref{solutionCG}) happens to be a good description of the dynamics of disk galaxies {\it regardless} of the variational setting. However, this good description
opens the issue of finding a setting for CG as a field theory other than the one currently used.

In conclusion, we pose a serious challenge to CG as a viable alternative theory of gravity capable of removing the requirement of dark matter to describe the phenomenology of cosmic structures.

\ack

 We sincerely thanks the referees for their very constructive comments that helped to clarify some issues of the original version of our manuscript. MC acknowledges support from the Grant no.~IDROL:51857 IDSIME:2016-0746 of Fondazione CRT. MC and AD also acknowledge partial support from the INFN grant InDark and Ministero dell’Istruzione, dell’Universit\`a e
della Ricerca: Departments of Excellence grant L.232/2016. LF acknowledges the INFN grant QGSKY, the local research project {\it  Metodi Geometrici in Fisica Matematica e Applicazioni (2015)} of Dipartimento di Matematica of University of Torino (Italy), and the grant INdAM-GNFM. This article is based upon work from COST Action (CA15117 CANTATA), supported by COST (European Cooperation in Science and Technology).

\appendix
\setcounter{section}{1}
\section*{Appendix: Gauge natural formalism reviewed}

CG is considered as a field theory where fields are sections of the configuration bundle $\calC$ with coordinates $(x^\m(x),g_{\m\n}(x))$, where the dynamics is covariant with respect to gauge transformations defined as automorphisms of $\calP$. $\calP=(P, M, p, G)$ is  a principal bundle, where $M$ is the manifold with coordinates
$x^\m(x)$, $G=(\R,+)$ is the Lie group of the set of real numbers with addition, $P$ is the set of all coordinates $(x^\mu, l)$, with $l$ the Lie group parameter, 
and $p$ is the map projecting the space $P\times M$ into $M$, i.e.~$p: P\times M \to  M:(l,x) \to x$ (Figure \ref{fig:sketch}).

Since $P$ is principal, its transition functions $x'^\mu(x)$ and $\om(x)$ are in the form
\begin{equation}
\begin{cases}
x'^\mu=x'^\mu(x)\\
l'=\om(x)+ l \; .\\
\end{cases}
\label{eq:GNtransf}
\end{equation}
Since these transformations also are affine, the bundle $P$ is, at the same time, principal and affine. As an affine bundle, it allows global sections. Based on our choice $(P, M, p, \R)$, $P=M\times \R$ is principal and affine, hence trivial.

In addition,  we can define the frame bundle $L(M)=(LM, M, \pi, \GL(m))$  for any manifold $M$, where $LM$ is the set of all pairs $(x,e_a$) with $e_a$ any basis of the tangent space to $M$ at the point $x \in M$; $\GL(m)$, namely the general linear group of degree $m$, is the set of invertible matrices $m\times m$, with $m$ the dimension of the manifold $M$; $\pi$ is the projection $\pi:LM \arr M :(x,e_a)\to x$. In other words, $L(M)$ is the principal bundle of  bases of tangent vectors to $M$ in any point. The frame bundle $L(M)$ is a $\GL(m)$-principal bundle. An automorphism on $L(M)$ is given by
\begin{equation}
 \begin{cases}
x'^\mu=x'^\mu(x)\\
 l'= \om(x) + l\\
 e'^\mu_a = J^\mu_\nu(x) \> e_a^\nu\\
 \end{cases}
\label{GGT}\end{equation}
where $e_a^\mu$ is any basis of the tangent space and $J^\mu_\nu(x)=\partial x'^\mu/\partial x^\nu$ is the Jacobian. 
We can paste these two principal bundles $\calP$ and $L(M)$ together to define the structure bundle $LM\times P$ with the Lie group $\GL(m)\times \R$.

\begin{figure}
\begin{center}
\includegraphics[width=0.2\textwidth]{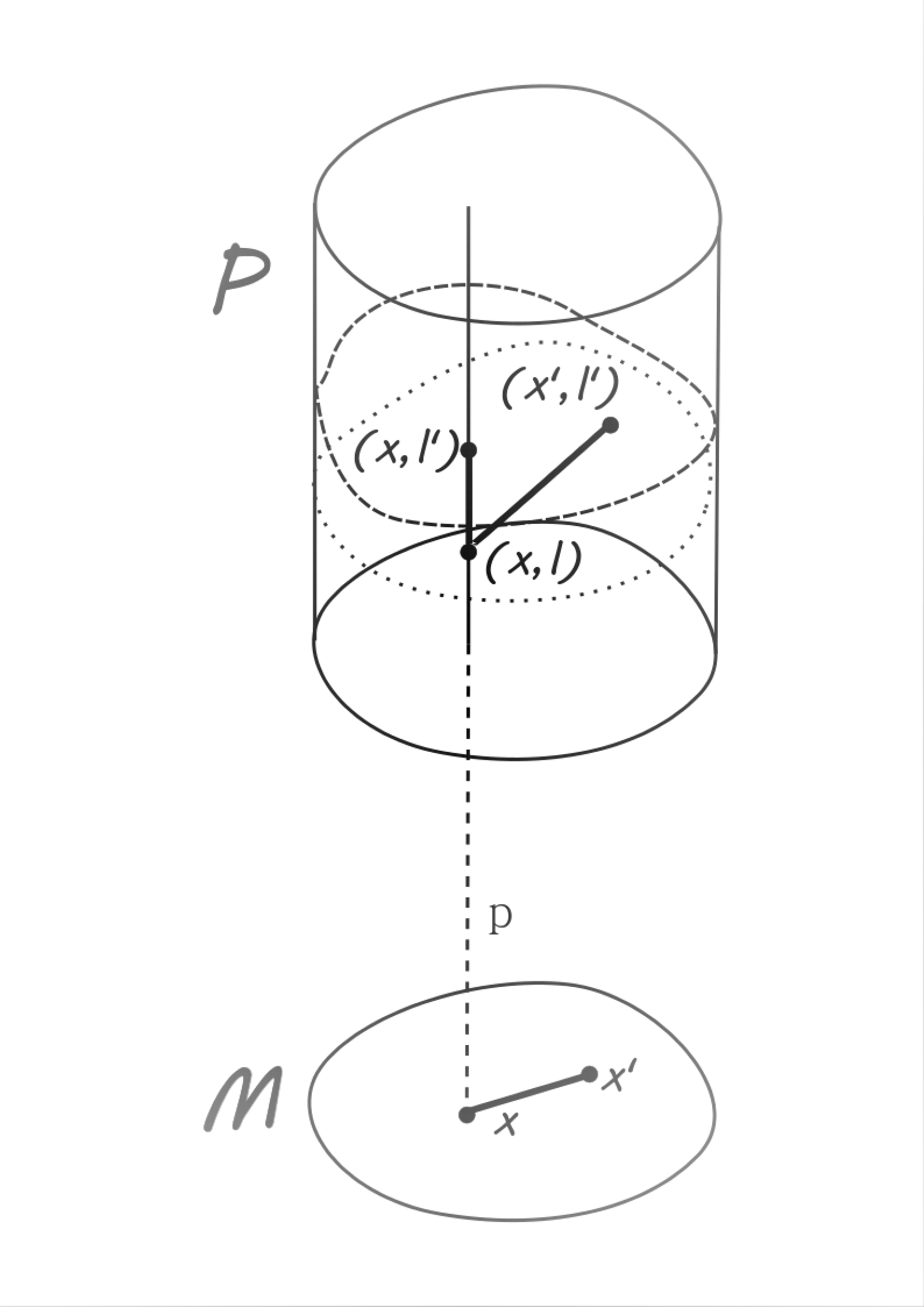}
\end{center}
\caption{\it The principal bundle $\calP$ and the manifold $M$. Two sections are shown. $p$ is the map projection that locally projects the pair of points $(x,l)$ into points $x$ on the spacetime manifold $M$. Pure gauge transformations are shown acting on points of a fiber as vertical automorphisms.} 
\label{fig:sketch}
\end{figure}

The configuration bundle $\calC$  is associated to the structure bundle $LM\times P$ by means of the action of the group $\GL(m)\times \R$ on $B(\eta)$
\beq
\la: \GL(m)\times \R\times B(\eta)\arr B(\eta)
\label{eq:lambda}
\eeq
where $B(\eta)$  denotes the set of all symmetric, non-degenerate, bilinear forms of Lorentzian signature $\eta=(3,1)$; the set $B(\eta)$ is an open set in the vector space of symmetric tensors of rank two  $S_2(\R^m) \simeq \R^{\frac{m(m+1)}{2}}$, parametrized by coordinates $g_{ab}$. In terms
of coordinates in each space, equation \eqref{eq:lambda} above reads
\begin{equation}
\la: (J_a^c, \om, g)\mapsto g'_{ab}=e^\om \bar J_a^c g_{cd} \bar J^d_b \; .
\end{equation}
The configuration bundle is then defined as $\calC=(L(M)\times \calP)\times_\la B(\eta)$, where $\times_\la$ indicates that the product of the spaces is
obtained through the action $\la$.
Points in $\calC$ are orbits $[e_a, x, l, g_{ab}]_\la$, namely the equivalence class of points related through the action $\la$. We can always choose the representative orbit $(\mathbb{I},x^\mu,0,g_{\mu\nu})$ in order to reduce ourselves to the coordinates $(x^\mu, g_{\mu\nu})$.

Local coordinates on $\calC$ are $(x^\mu, g_{\mu\nu})$ and they transform as
\begin{equation}
 \begin{cases}
x'^\mu=x'^\mu(x)\\
 g'_{\mu\nu}= e^{\om(x)}  J_\mu^\rho g_{\rho\si} \bar J^\si_\nu \; .\\
 \end{cases}
\label{GGTC}
\end{equation}

In this way, any automorphism on $\calP$ induces an automorphism on $\calC$. Hence the configuration bundle $\calC$ comes with a selected subgroup of transformations  $\Aut(\calP)\subset \Aut(\calC)$ which preserves the dynamics of the theory. Such transformations are called {\it generalized gauge transformations}.
The group of vertical automorphisms on $\calP$, i.e. $(x^\m,l)\to (x^\m,l')$, which leaves the coordinates $x^\m$ unaffected (see Figure \ref{fig:sketch}), is denoted by $\Aut_V(\calP)\subset \Aut(\calP)$. It also induces automorphisms on the configuration bundle, because $\Aut_V(\calP)\subset \Aut(\calP)\subset \Aut(\calC)$. These transformations of fields are called {\it pure gauge transformations}.

In gauge natural theories, spacetime diffeomorphisms and gauge transformations are completely unrelated. $\calP$ contains extra information that is not contained in the spacetime manifold $M$ and can be measured by any observer.
Gauge transformations canonically act on the associated configuration bundle $\calC$. Moreover, all sections of $\calC$, namely all fields, are required to be dynamical.

In natural theories, however, the entire information of the symmetries of the theory is encoded in the diffeomorphisms of the spacetime. There is no gauge symmetry. The structure bundle is then a natural bundle and thus all diffeomorphisms can be canonically lifted to $\Aut(\calC)$ \cite{Campigotto2015}. Campigotto and Fatibene \cite{Campigotto2015} showed that CG can also be described in the framework of natural theories with a specific choice of diffeomorphisms, $\om= \ln J$.

Let us define the associated bundle $\hat P= LM \times \R$ which, by construction, has fibered coordinates $(x^\mu, l)$ which transform as
\begin{equation}
\begin{cases}
x'^\mu=x'^\mu(x)\\
l'= \ln J + l \; .\\
\end{cases}
\end{equation}
We can see that the element $\ln J$ of group $(\R,+)$ acts by (left) translations onto $l$ so that the bundle $\hat P$ is by construction principal with the group $(\R,+)$. In this second case, unlike equation \eqref{eq:GNtransf} for the gauge natural description, the spacetime diffeomorphisms and the gauge transformations are not unrelated.  The consequences of these two different approaches in CG are illustrated in \cite{Campigotto2015}.

\section*{Refereces}

\begingroup\raggedright

\endgroup

\end{document}